\documentclass[11pt]{article}

\usepackage[utf8]{inputenc}
\usepackage[T1]{fontenc}
\usepackage{lmodern}
\usepackage[margin=1in]{geometry}
\usepackage[numbers,sort&compress]{natbib}
\usepackage[hidelinks]{hyperref}
\usepackage{microtype}
\usepackage{setspace}
\usepackage{parskip}
\usepackage{titlesec}
\usepackage{graphicx}
\usepackage{amsmath}
\usepackage{xurl}
\usepackage{hyperref}
\titleformat{\section}{\large\bfseries}{\thesection.}{0.5em}{}
\newenvironment{revblock}{}{}

\title{\textbf{Cycle-resolved explainability of energy storage impacts on whole-system cost and emissions}}
\author{
Zenghui Zhang$^{1,2}$ \and Wei He$^{2,*}$\\[0.5em]
\small $^1$School of Management, Hefei University of Technology, Hefei 230009, China\\
\small $^2$Department of Engineering, King's College London, London WC2R 2LS, United Kingdom\\
\small $^*$Corresponding author: \href{mailto:wei.4.he@kcl.ac.uk}{wei.4.he@kcl.ac.uk}
}
\date{}

\begin{document}

\maketitle

\begin{abstract}
Grid-scale storage is not intrinsically cost-reducing or emissions-reducing: whether a storage cycle lowers or raises whole-system cost and CO$_2$ depends on when, where, how long, what charges it, and what it displaces. Yet existing studies explain storage value either in aggregate at the system level or through asset-level market response, leaving a blind spot between whole-system value and the operational actions that create it. Here we resolve this blind spot by attributing whole-system cost and CO$_2$ consequences to individual charge--discharge cycles across present-day (2023) and planned (2030) UK and EU power systems. We show that aggregate indicators conceal sharply different cycle-level outcomes: in today's systems, commercially rational battery dispatch can reduce cost while increasing emissions, whereas long-duration energy storage (LDES) more consistently captures infrequent cross-timescale co-benefit cycles that jointly reduce both. Changing the dispatch objective redistributes storage within the existing cost--emission opportunity set, but supply-side decarbonisation shifts that opportunity set itself, expanding co-benefit opportunities across technologies. This complementary structure --- batteries serving recurrent opportunities, LDES capturing stress-period co-benefits --- persists robustly across wide variation in technology parameters, is driven by duration-gated access to distinct system states, and extends across technology cost and network location, identifying which capital pathways and which regions can most efficiently realise these system-valued cycles. The findings reveal three gaps between current market arrangements and whole-system value: an operational alignment gap, where better incentives could steer storage toward existing co-benefit cycles; a realisation gap, where long-duration co-benefit potential goes unexploited under current institutions; and a structural gap, where the available stock of co-benefit cycles remains limited until the supply mix itself decarbonises. By establishing a public-value benchmark independent of current market rules, this framework provides a basis for storage portfolio design, siting, and carbon-sensitive market reform that targets the operating regimes through which storage creates public value.
\end{abstract}

\section*{Introduction}
\begin{revblock}
Energy storage is central to renewable-based power-system decarbonisation: it helps integrate variable generation, supports security of supply under increasingly variable net demand, and can reduce system cost when deployed and operated well \cite{DeSisternes2016,Sepulveda2021}. Existing research explains this role from two main perspectives. Energy-system planning studies evaluate storage within least-cost or cost--emission optimisation \cite{Qin2023} and show why particular storage portfolios emerge in decarbonisation pathways \cite{Sepulveda2021,Dowling2020}, highlighting the roles of duration, clean-supply expansion, transmission and complementary flexibility resources in shaping system value \cite{Mantegna2024,Liu2023}. Asset-level studies explain how individual storage plants are sized and operated under prevailing price signals and market rules \cite{Olsen2019,Fares2017}, clarifying how arbitrage and other revenue streams shape dispatch and project economics. These literatures therefore provide two important but separate forms of explainability: \emph{planning explainability}, which explains why storage matters for system design, and \emph{asset-level explainability}, which explains how storage responds to available market incentives. What remains missing is \emph{system-to-operation explainability}: the ability to translate whole-system value into interpretable operating opportunities and actions across technologies, durations and network locations \cite{Qiu2024,Brinkerink2024}.

This missing layer matters because storage is not intrinsically cost-reducing or emissions-reducing \cite{Beuse2021,Arbabzadeh2019}. Power systems contain both co-benefit opportunities, where charging and discharging jointly lower cost and CO$_2$ \cite{Beuse2021,Olsen2019}, and trade-off opportunities, where one improves at the expense of the other. Which opportunities are available depends on when storage charges and discharges, how long energy is shifted, which generators supply charging energy, which resources are displaced at discharge, and where these interactions occur in the network \cite{Beuse2021,Sepulveda2021}. We can therefore know that storage is valuable in aggregate from system studies, and measure plant-level performance under prevailing market rules, yet still be unable to specify which cycle-to-cycle actions, at what times, in which locations, and with which storage technologies create public value, and whether public values of storage are captured by existing market mechanisms \cite{Qin2023,Kucuksayacigil2025}. That weakens the evidence base for decarbonisation policy, storage remuneration and market design, because it blurs the distinction between operation that lowers whole-system cost and CO$_2$ and operation that merely follows private price opportunities.

To bridge this gap from whole-system values to storage operations, the challenge is structural. Generators contribute by injecting electricity at a given time and place; storage contributes by withdrawing electricity under one set of conditions and re-injecting it under another. Its system impact therefore arises from the pairing of charge and discharge across time and space, together with the charging source, the generation displaced at discharge, interregional exchange, and technology characteristics such as duration and round-trip efficiency \cite{Beuse2021,Sepulveda2021}. These interactions create heterogeneous system impacts even within a single asset in day-to-day operation, and they change with system conditions. Aggregate indicators such as annual values, average emission factors, marginal emission factors and shadow prices are useful summaries of system conditions, but they do not explain which individual storage actions create those outcomes \cite{SilerEvans2012,Elenes2022}. In other words, even if two storage systems have similar aggregate indicators, they may inherently operate with completely different operational patterns and contribute differently to the system. Conversely, asset-level analyses explain price response and profitability, but often leave public system consequences implicit or unmeasured, as they only capture the mechanisms available in today's markets, which are designed primarily for conventional generators \cite{Qin2023}. The result is a persistent blind spot between system value and operational action.

This gap is not merely analytical but institutional. Current electricity markets are designed around conventional generators, so storage participates by responding to price signals as an individual asset and whole-system consequences are passively received rather than governed \cite{Qin2023}. Efforts to develop storage-specific flexibility products and market mechanisms are underway in many jurisdictions, but they face a common challenge: the operating regimes most valuable to the system have not yet been identified or specified, making it difficult to design procurement or remuneration around them \cite{DESNZ2024REMA,NaviaSimon2025}.

We address this blind spot by making whole-system value legible at the level where storage actually acts: the individual charge--discharge cycle. Because each cycle's consequences depend on timing, duration, location, charging source and displaced generation \cite{Beuse2021,Sepulveda2021}, we develop a cycle-resolved operation-to-system attribution and explainability framework that assigns whole-system cost and CO$_2$ consequences to individual cycles and maps them into a four-quadrant cost-emission space. Applied to present-day (2023) and near-future (2030) UK \cite{NESO2024} and EU power systems, the framework explains a common opportunity structure: today’s systems contain both co-benefit and trade-off cycles \cite{Beuse2021}, co-benefit opportunities remain limited, long-duration storage is better positioned to capture infrequent cross-timescale co-benefit cycles \cite{Sepulveda2021,Staadecker2024}, and short-duration batteries complement it through more frequent cycling across the remaining opportunity set. We further show that changing the dispatch objective mainly redistributes storage within the existing opportunity set, whereas supply-side decarbonization shifts that opportunity set itself. Because it quantifies public value independently of current market rules, the framework establishes a benchmark against which actual market outcomes can be assessed --- identifying which cycle-level values existing arrangements already capture and which they miss. We then extend the same framework across technology cost and regional exchange, providing a basis for storage portfolio design, siting and market signals that reward flexibility according to whole-system value. We focus on grid-connected, post-generation storage, where electricity is withdrawn from and re-injected into the power system. Fuel-side or pre-generation storage can provide complementary flexibility, but comparing these options requires a different system boundary and is outside the scope of this study.
\end{revblock}

\section*{Cycle-resolved cost--emission attribution makes storage operation explainable}
The individual storage cycle is the explanatory unit of the framework. For each charge--discharge cycle, we compute a cycle emission factor (CEF) and a cycle cost factor (CCF), which quantify the whole-system CO$_2$ and cost consequences per unit of discharged energy relative to the resources used for charging and the residual generation displaced during discharge. Unlike average or marginal indicators, these metrics attribute system outcomes to the paired operating action that creates them \cite{SilerEvans2012,Elenes2022}. Formal definitions are given in Methods (Eqs.~\eqref{eq:cef}--\eqref{eq:agg-discharging-emission-factor}), and aggregated statistics are reported in Table~S1.

In the baseline cases, batteries are represented as stylized short-duration storage with a round-trip efficiency of 0.9 and an end-of-day discharge constraint, whereas LDES is represented as lower-efficiency storage with a round-trip efficiency of 0.7 and no daily cycling constraint. These assumptions separate recurrent short-duration cycling from cross-timescale storage operation and are later relaxed through parameter sweeps over duration, efficiency and rated power.

\begin{figure}[!htbp]
    \centering
    \includegraphics[width=\linewidth]{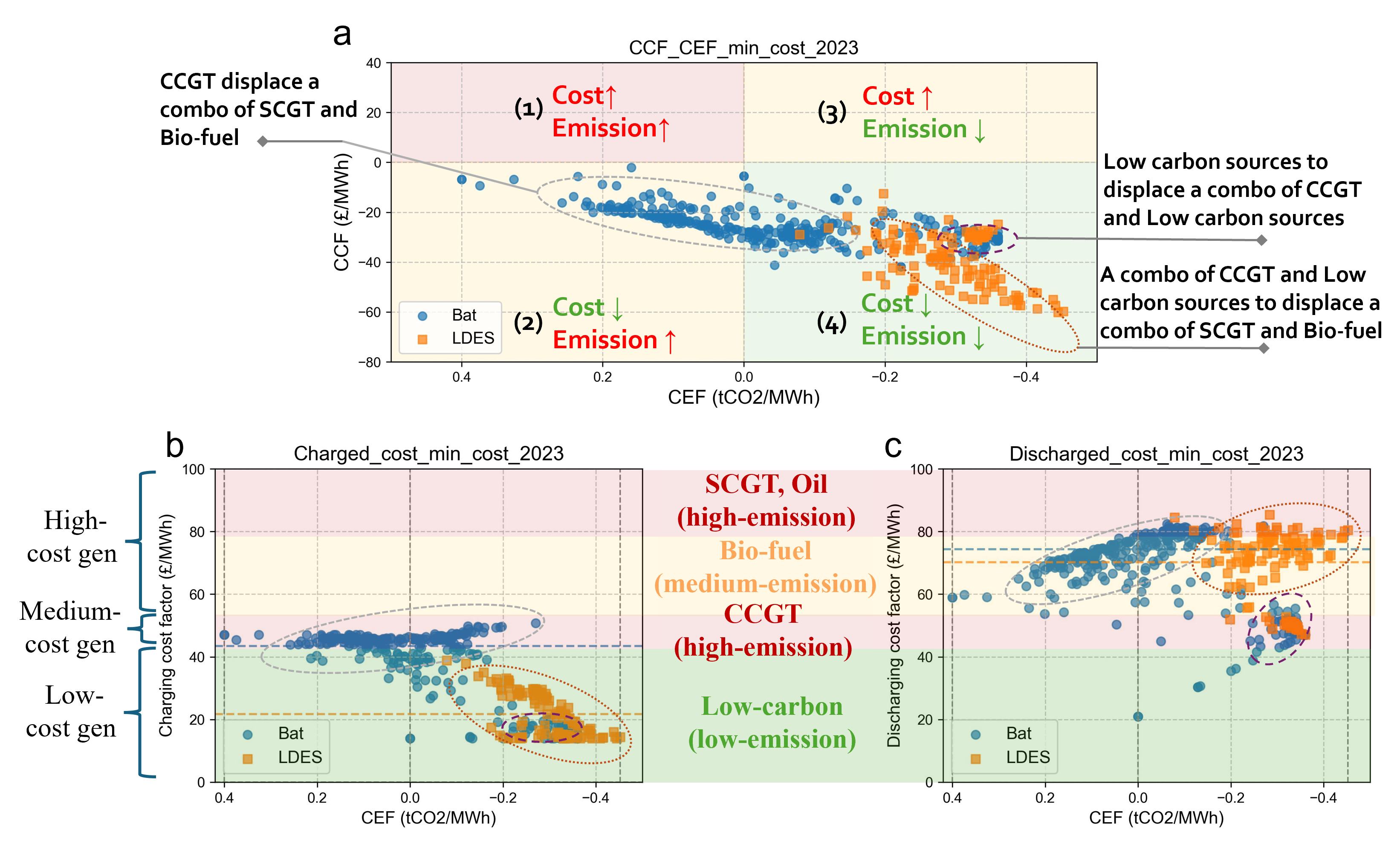}
    \caption{\textbf{Multi-cycle distribution of energy storage across four cost--emission quadrants and associated cost decomposition.}
    \textbf{a,} Distribution of storage operation cycles under the \textit{UK\_min\_cost\_2023} scenario. Each point represents one symmetric charge–discharge cycle extracted from regionally aggregated storage operation. Negative CEF or CCF values indicate reductions in whole-system emissions or cost, respectively. The four shaded quadrants classify cycles into co-benefit, trade-off, or misaligned regimes. An annotation “A displaces B” means that storage charged from generation mix A and displaced generation mix B during discharge.
    \textbf{b,c,} Charging and discharging cost-factor decomposition for the same cycles. Horizontal dashed lines show aggregated values by storage technology. Generator categories are grouped by relative cost and carbon intensity, with definitions provided in Methods. The CEF axis is reversed so that stronger emission reductions appear to the right.}
    \label{fig:fig1}
\end{figure}

Plotting all cycles in the joint CEF--CCF plane reveals the system’s opportunity structure: a distribution of operationally interpretable outcomes rather than a single annual effect (Fig.~\ref{fig:fig1}a). The sign combination defines four quadrants. Quadrant~4 contains co-benefit cycles that reduce both cost and emissions. Quadrants~2 and~3 contain trade-off cycles that improve one objective while worsening the other. Quadrant~1 contains cycles that worsen both in the cost--CO$_2$ plane, although some may still arise when operational constraints limit access to cleaner or cheaper alternatives. The four-quadrant map is the ``semantic" layer of the explainability framework: it distinguishes which storage actions create co-benefits, which create trade-offs, and which remain necessary but misaligned with cost or carbon objectives.

The baseline \textit{UK\_min\_cost\_2023} case makes this opportunity structure visible. Even under a single dispatch objective, storage operates across both co-benefit and trade-off opportunities \cite{Beuse2021,Hittinger2015}. Most cycles lie in quadrant~4, but batteries retain a substantial quadrant-2 distribution, with 42\% of battery cycles reducing cost while increasing emissions, whereas LDES is concentrated more strongly in quadrant~4 (Fig.~\ref{fig:fig1}a). Yet both technologies still appear beneficial on average, with negative aggregated CEFs of $-0.02$ tCO$_2$/MWh for batteries and $-0.29$ tCO$_2$/MWh for LDES (Table~S1). Aggregate metrics therefore conceal a key distinction: similar annual net benefits can arise from very different mixtures of co-benefit and trade-off cycles.

The charging--discharging decomposition in Fig.~\ref{fig:fig1}b,c explains these positions in the cost--emission plane. Cost-reducing cycles arise when storage charges from a lower-cost mix and displaces higher-cost residual generation at discharge. In the baseline case, the aggregated discharging cost factor exceeds the aggregated charging cost factor for both technologies, confirming that system benefit is created by charge--discharge pairing rather than by charging or discharging in isolation. The stronger concentration of LDES in quadrant~4 therefore reflects more favourable temporal pairing, not a uniformly different average condition. This is consistent with duration allowing LDES to access less frequent system states in which clean-surplus charging and stressed-period discharge jointly create co-benefits \cite{Sepulveda2021,Dowling2020}, while batteries remain more exposed to the present-day trade-off frontier.

CEF and CCF provide the attribution explainable layer of the framework, and the four-quadrant map provides its semantic explainable layer. Together, they show that present-day systems contain both co-benefit and trade-off opportunities, and that batteries and LDES occupy them differently even when annual averages suggest similar net value. System-to-operation explainability therefore does not simply disaggregate storage value. It reveals which operating opportunities exist, which technologies capture them, and where misalignment between private dispatch and public value begins.

\section*{Distributional shifts distinguish operational retuning from structural change}
\begin{figure}[!htbp]
    \centering
    \includegraphics[width=\linewidth]{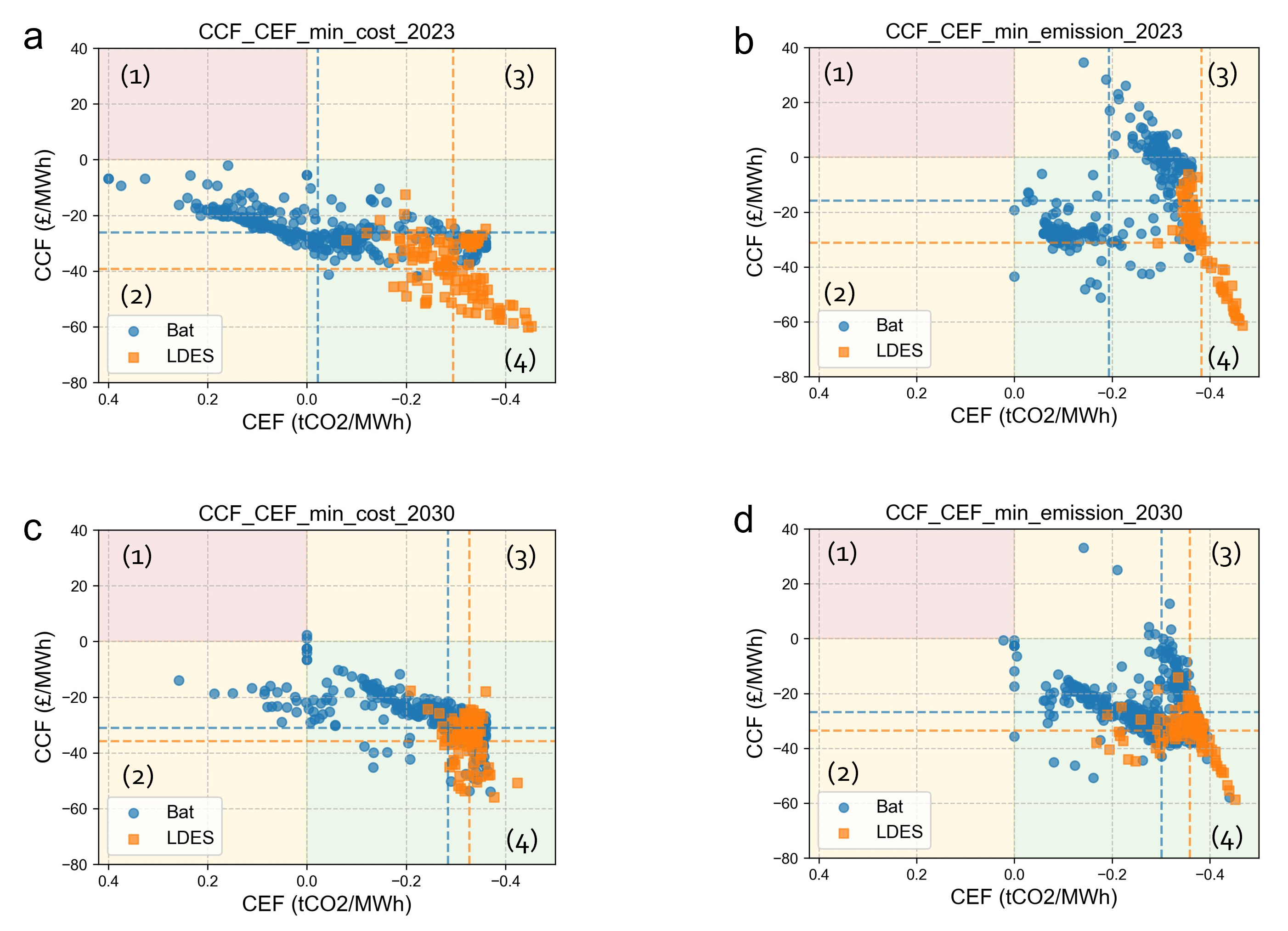}
    \caption{\textbf{Effects of system evolution and operational priorities on storage cycle cost--emission outcomes.} Subplots show the distribution of storage cycles under \textbf{a,} cost minimization in System 2023, \textbf{b,} emission minimization in System 2023, \textbf{c,} cost minimization in System 2030 and \textbf{d,} emission minimization in System 2030. Each panel shows cycle outcomes under one system–objective combination; changes are inferred by comparing distributions across panels. Horizontal and vertical dashed lines denote the aggregated cycle cost factor and aggregated cycle emission factor across all cycles, respectively. Colours indicate storage technologies as defined in the legend, and the quadrant labels and background shading are consistent with Fig.~\ref{fig:fig1}a.}
    \label{fig:fig2}
\end{figure}

The quadrant distribution distinguishes two ways storage value changes: \emph{operational retuning} and \emph{structural system change}. Fig.~\ref{fig:fig2} compares cost-minimizing and emission-minimizing dispatch in the 2023 and 2030 systems. Viewed through the explainability framework, these comparisons show whether storage is redistributed within an existing cost--emission opportunity set or whether that opportunity set itself shifts.

In the 2023 system, where co-benefit opportunities remain limited, switching the objective from cost minimization to emissions minimization mainly retunes the distribution of residual trade-off cycles while leaving quadrant~4 dominant (Fig.~\ref{fig:fig2}). For batteries, the shift removes quadrant~2 almost entirely and introduces a substantial quadrant~3 share, making the aggregated CEF more negative while weakening cost savings. Specifically, the aggregated battery CEF shifts from $-0.02$ to $-0.19$ tCO$_2$/MWh, while the aggregated CCF becomes less negative, moving from $-26.10$ to $-15.84$ \pounds/MWh (Table~S1). By contrast, for LDES in the UK cases, LDES remains entirely within quadrant~4 under both objectives, so retuning changes the density and aggregate position of co-benefit cycles rather than moving LDES across quadrants. Its aggregated CEF still becomes more negative, by 0.09 tCO$_2$/MWh, but this gain is accompanied by a smaller 8.06 \pounds/MWh weakening in the aggregated CCF. Operational retuning in the current system therefore creates additional abatement mainly by steering marginal battery cycles away from cost-led trade-off regimes, while LDES already operates close to a co-benefit regime. Retuning therefore changes how storage selects from the existing opportunity set, but does not materially alter that opportunity set itself.

System decarbonisation produces a qualitatively different shift. Moving from the 2023 to the 2030 system pushes the overall cycle distribution toward quadrant~4, so that cost-optimal operation increasingly coincides with emissions reduction rather than trading against it. This is most visible for batteries: under cost minimization, the number of quadrant-4 cycles doubles from 202 to 405, while both the aggregated CEF and aggregated CCF improve (Fig.~\ref{fig:fig2}a,c; Tables~S1 and S2). In the UK, LDES remains dominated by quadrant~4 throughout, indicating that a cleaner supply mix does not so much redirect its operating regime as deepen the availability of low-cost, low-emission cross-timescale opportunities. The EU cases show the same broad tendency toward greater quadrant-4 prevalence, albeit with greater heterogeneity across systems and objectives (see Supplementary Fig.~S1; Table S3). A cleaner supply mix therefore expands the set of storage actions that are intrinsically aligned with both cost and emissions objectives \cite{He2020,Moradi2024}, and systematically moves battery operation away from the present-day trade-off frontier.

Operational retuning and structural system change affect storage operation in different ways. Retuning changes how storage selects from the existing cost--emission opportunity set: in the current system, batteries move mainly along the trade-off frontier, whereas LDES is retuned largely within an already co-benefit regime. Structural decarbonisation changes the opportunity set itself by expanding the co-benefit region and relaxing the trade-offs that constrain present-day operation, thereby allowing cost-optimal battery cycles to move from trade-off toward co-benefit operation. This distinction is the practical meaning of system-to-operation explainability. It shows how much value can be captured within the current system through better operational choices, and how much depends on deeper change in the surrounding supply mix. It also clarifies why batteries and LDES retain complementary roles: batteries remain more exposed to trade-off operation in present-day systems, whereas LDES is better positioned to capture the less frequent co-benefit opportunities that persist across system states \cite{Sepulveda2021,Staadecker2024}. For storage operation, the framework identifies which cycles should be encouraged, avoided or deprioritized if operational incentives are to align storage dispatch with whole-system value. For market-design discussions, it helps distinguish improvements that may be delivered through better operating incentives from those that require broader system decarbonisation to make co-benefit cycles more abundant \cite{Qin2023,DESNZ2024REMA}.

\section*{System-to-operation explainability reveals complementary operating regimes of batteries and LDES}
The preceding analysis shows that batteries and LDES occupy the opportunity structure differently, but this finding rests on a single baseline parameterisation. We next test whether the complementarity persists across the broader performance ranges spanned by major storage technologies. We therefore sample round-trip efficiency, rated power and maximum discharge duration over a broad configuration space and re-optimise the system for each configuration (see Supplementary Note~S1). Fig.~\ref{fig:fig3} tests complementarity from three perspectives: operating intensity (Fig.~\ref{fig:fig3}a,b), contribution per unit capacity (Fig.~\ref{fig:fig3}c,d), and carbon-abatement timing during high-emission system states (Fig.~\ref{fig:fig3}e,f), and Supplementary Figs.~S2 and S3 provide the corresponding cycle-level interpretation.

\begin{figure}[!htbp]
    \centering
    \includegraphics[width=\linewidth]{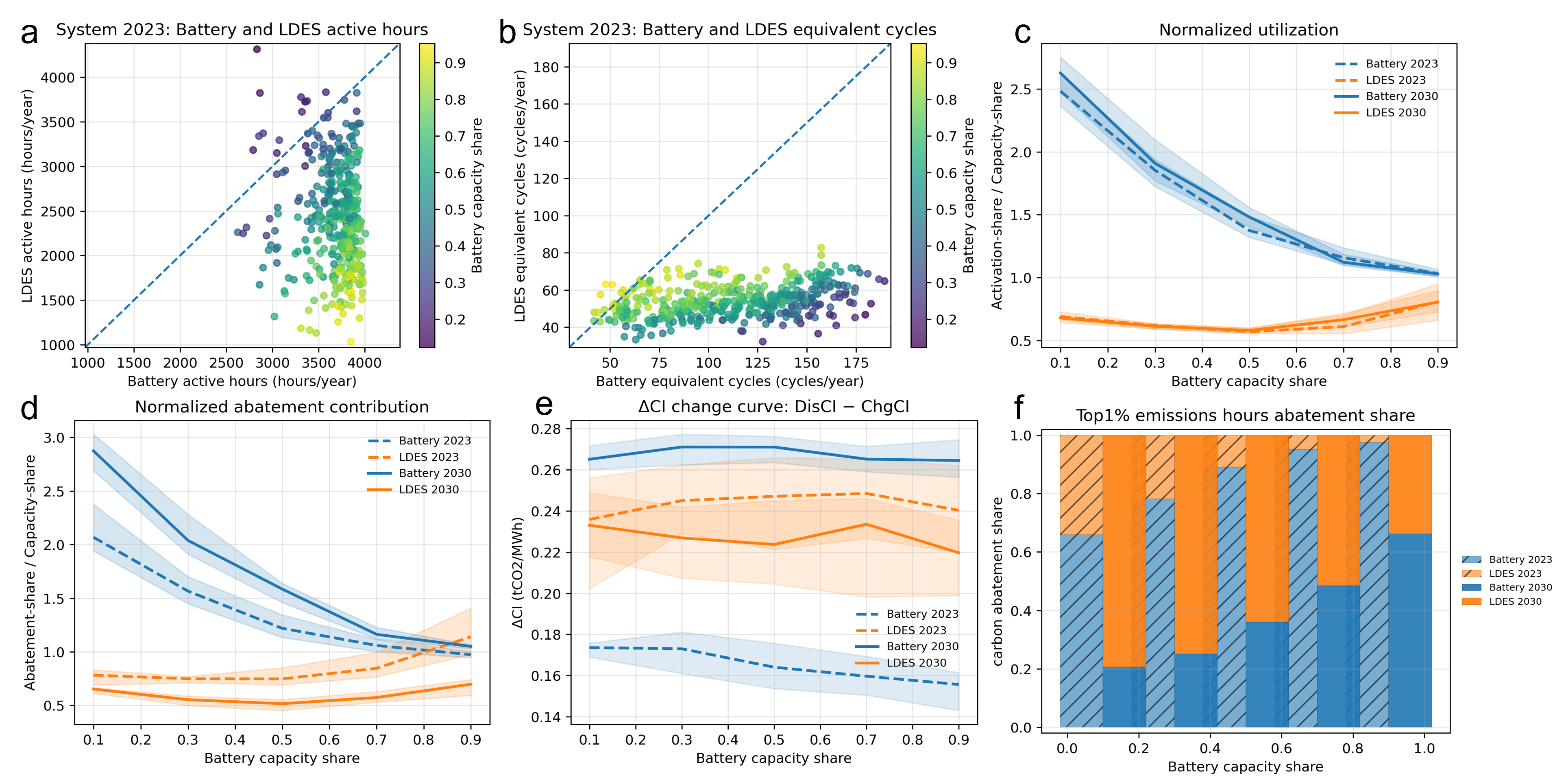}
    \caption{\textbf{Robust complementary operational roles of battery and LDES across the storage parameter space.} 
    \textbf{a,} Activated hours of battery versus LDES in System 2023, with each point representing one scenario and colour indicating the battery capacity share (dashed line, $y=x$). 
    \textbf{b,} Equivalent cycle numbers of battery versus LDES in System 2023, defined as annual activated energy divided by installed capacity; colour coding and the reference line are as in \textbf{a}. 
    \textbf{c,} Normalized utilization rate, defined as activated energy share divided by capacity share, for both storage types as a function of battery capacity share in System 2023 (dashed lines) and System 2030 (solid lines), with shaded uncertainty ranges. 
    \textbf{d,} Normalized emission-reduction contribution, defined as emission-reduction share divided by capacity share, for both storage types versus battery capacity share, with line styles and shading as in \textbf{c}. 
    \textbf{e,} Charging and discharging carbon intensities, where ChgCI and DisCI denote the total charging-related emissions and total discharging-related emission reductions per unit of charged energy, respectively; curves show $\Delta$CI = DisCI $-$ ChgCI as a function of battery capacity share. 
    \textbf{f,} Share of storage-induced emission reductions occurring during the top 1\% highest-emission hours of the year, with hatched and solid bars corresponding to System 2023 and System 2030, respectively.}
    \label{fig:fig3}
\end{figure}

Across this ensemble, batteries and LDES retain a robust complementarity in how they capture system value \cite{Sepulveda2021,Staadecker2024}. Batteries remain active across a relatively stable range of hours but vary substantially in equivalent cycle count, indicating that design variation mainly changes how intensively they exploit recurrent operating opportunities (Fig.~\ref{fig:fig3}a,b). LDES shows the opposite pattern: active hours vary more widely, but equivalent cycles remain comparatively stable, implying that design variation mainly changes when long-duration storage is called upon rather than the depth of use once activated. In the language of the opportunity framework, batteries continue to serve the recurrent portion of the opportunity set, whereas LDES captures less frequent cross-timescale opportunities associated with system stress \cite{Sepulveda2021,Aspitarte2024}. At the cycle level, battery operation combines a quadrant-4-dominant structure with a persistent quadrant-3 contribution, so parameter changes mainly alter how intensively the recurring opportunity set is exploited. LDES, by contrast, is governed almost entirely by long quadrant-4 cycles, so parameter changes mainly alter the timing and duration of event-driven deployment rather than the character of the operating regime itself. The underlying mechanism is duration-gated access to different system states: short-duration storage accesses recurrent diurnal variation, whereas long-duration storage reaches infrequent multi-day stress events whose co-benefit value depends on sustained clean charging and prolonged displacement. Because these temporal regimes are structurally distinct in the system's net-demand profile, the complementarity is not an artefact of the binary technology labelling but emerges continuously across the duration parameter space. This complementarity persists from System~2023 to System~2030 (Fig.~\ref{fig:fig3}c; Fig.~S2).

System decarbonisation changes the balance of opportunities available to each technology, but not the underlying complementarity. In the cleaner 2030 system, batteries deliver substantially greater emissions reduction per unit capacity, with normalized abatement contribution rising to around 2.9, compared with about 2.0 in 2023 (Fig.~\ref{fig:fig3}d). This is consistent with the larger charging--discharging carbon-intensity gap in 2030 and, at the cycle level, with a stronger concentration of battery operation in quadrant~4 (Fig.~\ref{fig:fig3}e; Fig.~S3a). As cleaner supply expands, batteries therefore gain more routine co-benefit opportunities in day-to-day operation \cite{Luderer2022,Victoria2020}. At the same time, the residual system challenge becomes more concentrated in prolonged high-emission periods. Here LDES becomes more important: across the sampled battery-capacity shares, the battery share of storage-induced abatement during the top 1\% highest-emission hours falls from 78--97\% in 2023 to 21--48\% in 2030, while LDES sustains substantially longer abating cycles during these periods (Fig.~\ref{fig:fig3}f; Fig.~S3b,c). Decarbonisation therefore expands routine co-benefit cycling for batteries while sharpening the role of LDES in residual stress-period mitigation \cite{Sepulveda2021,Dowling2020}. This robustness confirms that complementary operating roles are not artefacts of baseline assumptions, but it leaves a practical question unanswered: which existing technology pathways can deliver these distinct cycle regimes, and at what capital cost? Without this link, the identified public value remains an operating-level insight that cannot yet inform procurement or portfolio investment.

\section*{Mapping system-valued storage cycles to technology-specific abatement costs}
\begin{figure}[!htbp]
    \centering
    \includegraphics[width=\linewidth]{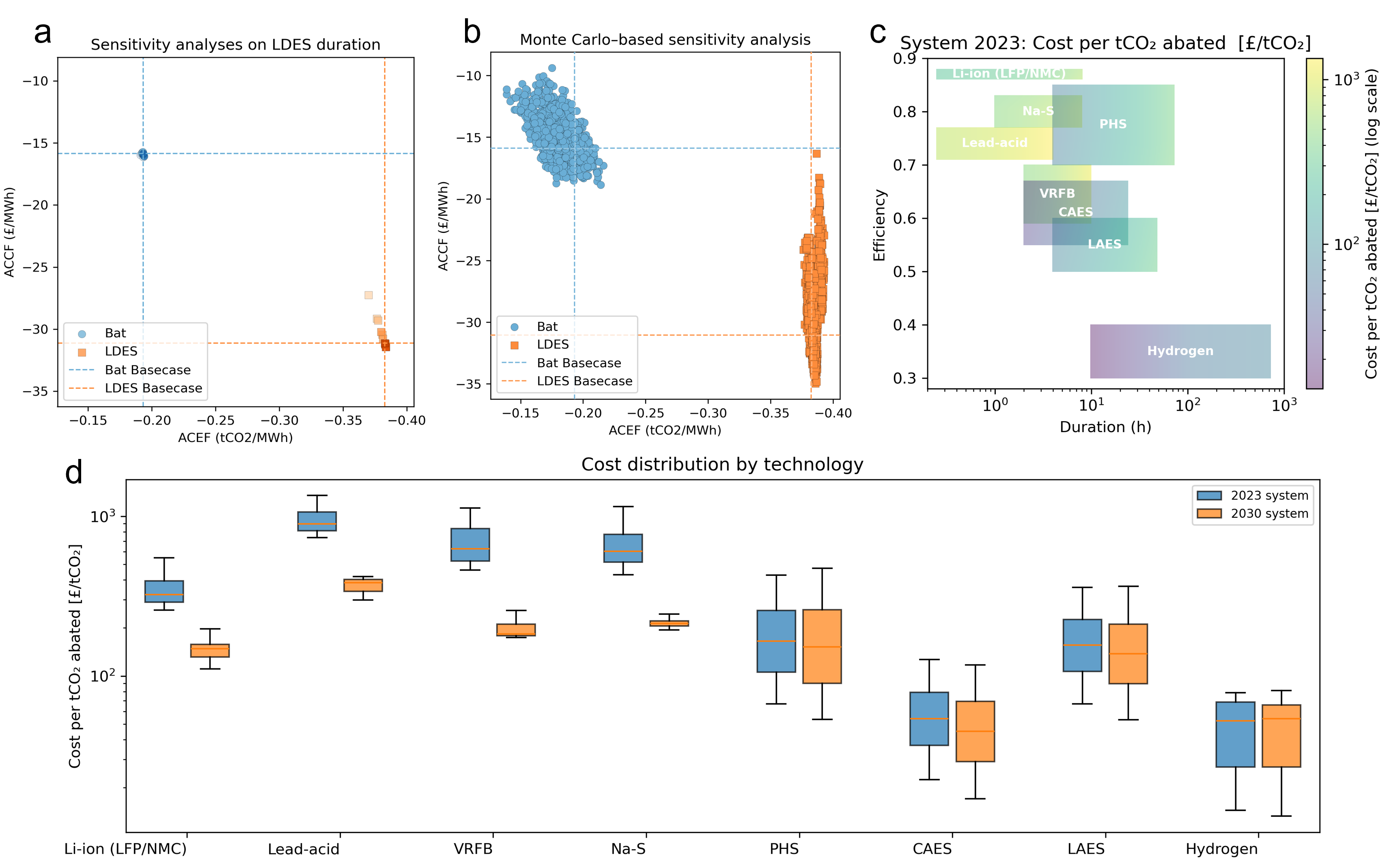}
    \caption{\textbf{Technology-specific annualized capital cost per tonne of operational CO$_2$ abatement.} 
    \textbf{a,} Aggregated CEF–CCF outcomes under LDES-duration sensitivity. Each scatter point represents one sensitivity run, with colour intensity indicating LDES duration (from light to dark: 1, 2, 3, 4, 5, 7, 14, 30, 61, 92, 183 and 365 days). Blue and orange points denote the aggregated CEF–CCF outcomes for batteries and LDES, respectively. Dashed lines labelled “base case” indicate the corresponding average aggregated values under the \textit{UK\_min\_emission\_2023} scenario, with colours matching the legend. 
    \textbf{b,} Aggregated CEF–CCF outcomes under Monte Carlo parameter sampling. 
    \textbf{c,} Distribution of technology-specific unit abatement costs in System 2023. Each rectangle represents a storage technology, with its horizontal and vertical spans corresponding to the typical ranges of duration and efficiency, respectively; rectangle colour encodes the unit abatement cost (£/tCO$_2$, logarithmic scale). 
    \textbf{d,} Comparison of unit abatement costs across systems. Each pair of boxplots corresponds to one storage technology and shows the distribution of unit abatement costs (£/tCO$_2$) under System 2023 and System 2030.}
    \label{fig:fig4}
\end{figure}

Having identified which storage cycles are aligned with whole-system cost and carbon objectives, we next ask which technology pathways can realize these cycle regimes, and at what capital cost. Using the parameter sweep above as the operational basis, we map cycle-resolved abatement opportunities onto literature-based ranges of duration, round-trip efficiency, installed cost and lifetime, and convert feasible operational abatement into annualized capital cost per tonne of CO$_2$ abated (\pounds/tCO$_2$, Supplementary Note~S2; Table S4) \cite{Aspitarte2024,Yuan2022}. This metric is not a full lifecycle cost or social cost of each technology. It is an operation-based abatement capital cost, defined as annualized storage capital cost divided by the annual operational CO$_2$ abatement attributed to storage cycles. Fig.~\ref{fig:fig4} summarizes this mapping for the UK in the emissions-minimizing 2023 system and shows how it shifts in System~2030.

The parameter sweep in Fig.~\ref{fig:fig4}a,b shows that the opportunity structure identified above is robust across the performance ranges spanned by major storage technologies: extending duration from day-scale to week-scale yields the largest gain in joint cost--emission performance, after which marginal gains diminish, and the broader ensemble confirms that this advantage persists across wide design variation (Fig.~S4; Table S5). This provides the bridge from operating roles to technology pathways.

Fig.~\ref{fig:fig4}c shows that the annualized capital cost of realizing storage abatement depends more strongly on feasible duration than on round-trip efficiency across realistic technology ranges \cite{Sepulveda2021,Aspitarte2024}. Short-duration batteries occupy a relatively narrow, higher-cost region because they access fewer of the infrequent co-benefit opportunities identified above and do so with higher capital cost per unit of realized abatement. Longer-duration pathways span a broader and generally lower-cost region because they are better aligned with the cross-timescale co-benefit cycles revealed by the explainability framework (Supplementary Fig.~S5a; Note~S3). Among short-duration options, lithium-ion batteries remain the most cost-effective \cite{Jiang2025a}, with a median operation-based abatement cost of \pounds323.35/tCO$_2$. Among longer-duration options, hydrogen-based pathways appear low-cost within the adopted literature cost range and system boundary \cite{Shafiee2024,Yuan2022}, with a median cost of \pounds52.34/tCO$_2$. These operation-based abatement costs exclude embodied emissions and life-cycle impacts, and LDES capital costs remain subject to significant uncertainty; the ordering should therefore be read as indicating which capital pathways are structurally favoured by the opportunity set, not as definitive technology rankings. More generally, technologies with similar quadrant exposure can still differ substantially in annualized abatement cost, because capital structure determines how cheaply operational value can be realized.

This ordering remains broadly stable as the system decarbonizes (Fig.~\ref{fig:fig4}d). In System~2030, both batteries and LDES shift further toward quadrant~4 exposure, indicating that cleaner supply improves the operating opportunity set across technologies (Fig.~S5b,c). However, this operating-side improvement is broadly technology-agnostic, whereas realized abatement cost remains shaped by capital-side differences in duration, cost and lifetime. Across representative LDES options, median abatement cost averages \pounds106.83/tCO$_2$ in 2023, 83\% lower than for batteries; in 2030, it falls to \pounds97.13/tCO$_2$ and remains 58\% lower. The gap narrows because batteries gain more routine co-benefit opportunities in the cleaner system, but it does not close. Technology choice is therefore not just a matter of performance specification, but of which capital pathway can most efficiently realize distinct co-benefit and trade-off-sensitive operating roles. The framework thus provides a basis for procurement, storage portfolio design and policy support that targets the delivery of whole-system-valued cycle regimes rather than treating storage technologies as interchangeable.

\section*{Spatial explainability shows how siting conditions and regional exchange shape storage value}
The preceding sections identify which operating regimes create public value and which technologies can deliver them, but treat the system as spatially uniform. In practice, the same cycle regime may produce co-benefits in one location and trade-offs in another, depending on the regional generation mix and interregional exchange. We therefore extend the explainability framework across space by resolving storage operation within a networked system. The aim is not to simulate physical power flows at nodal resolution, but to attribute effective charging sources and displaced generation under interregional exchange constraints. Using a local-first, clean-first accounting rule (Methods), we trace the effective generation mix serving regional demand and storage after interregional exchange (Figs.~S6,S7). This converts network interactions into region-specific charging and discharging conditions and shows how location reshapes trade-off and co-benefit opportunities.

\begin{figure}[!htbp]
    \centering
    \includegraphics[width=\linewidth]{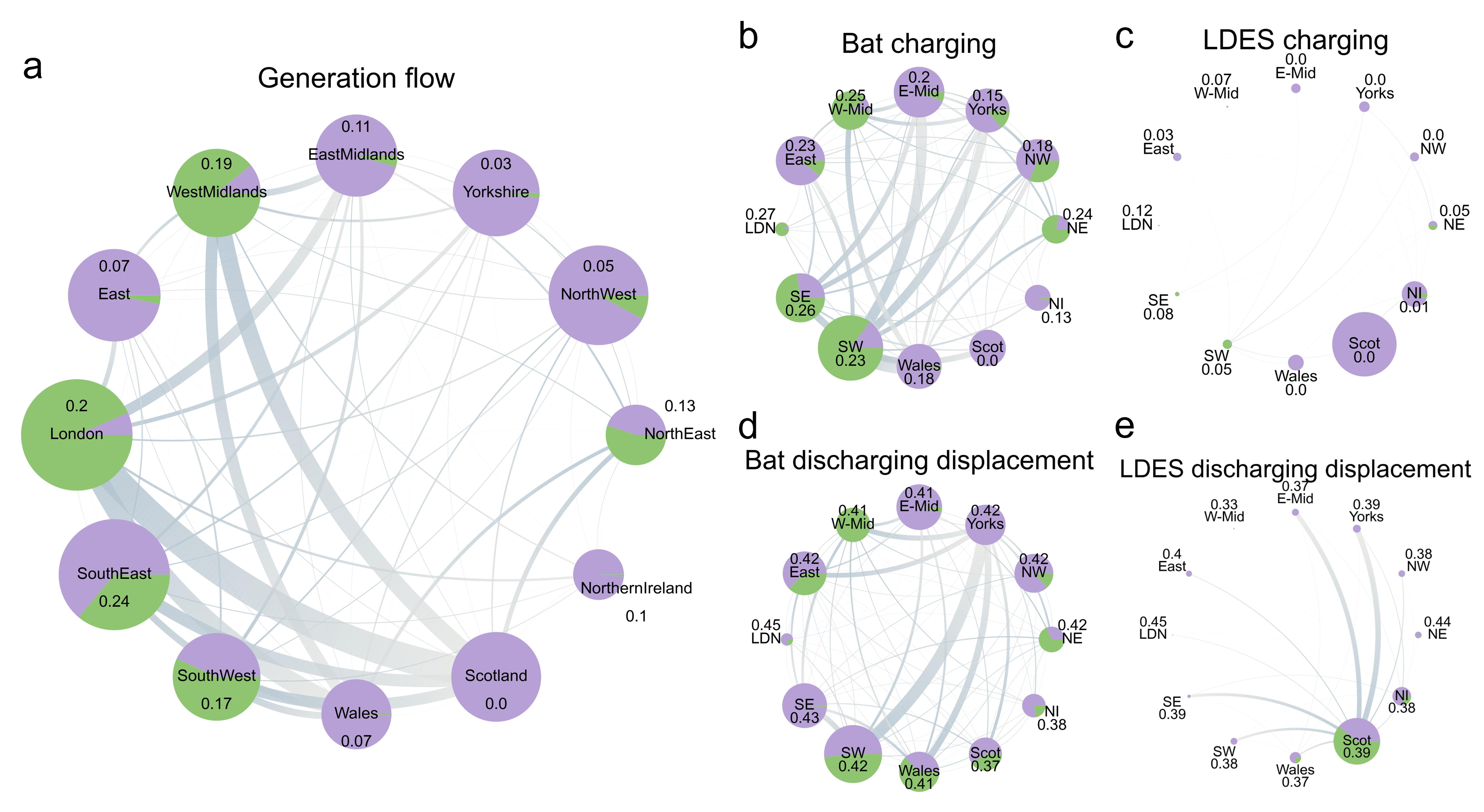}
    \caption{\textbf{Interregional electricity and storage-related energy exchanges.} 
    \textbf{a,} Interregional electricity supply and demand. Each node represents a UK region, with node size proportional to total electricity demand. The purple share indicates demand met by local generation, and the green share indicates imported electricity. Labels denote demand carbon intensity (tCO$_2$/MWh), defined as the average carbon intensity of all generation serving local demand. Lines represent interregional electricity flows, with thickness proportional to exchanged volume and colour gradients indicating flow direction from supplying regions to receiving regions. 
    \textbf{b,c,} Interregional charging of battery and LDES. Node size indicates total charging volume; purple and green shares denote charging supplied by local and external generation, respectively. Labels show the aggregated charging emission factor (tCO$_2$/MWh). Links in \textbf{b,c} represent accounting-based effective charging supply relationships rather than physical power flows.
    \textbf{d,e,} Interregional discharging of battery and LDES. Node size indicates total discharge volume; purple and green shares represent discharge displacing local and external generation, respectively. Labels show the aggregated discharging emission factor (tCO$_2$/MWh). In \textbf{d,e}, links represent displacement relationships rather than physical power flows. Subplots \textbf{b--e} use a common normalized scale to facilitate comparison across storage-related exchanges.}
    \label{fig:fig5}
\end{figure}

\begin{figure}[!htbp]
    \centering
    \includegraphics[width=\linewidth]{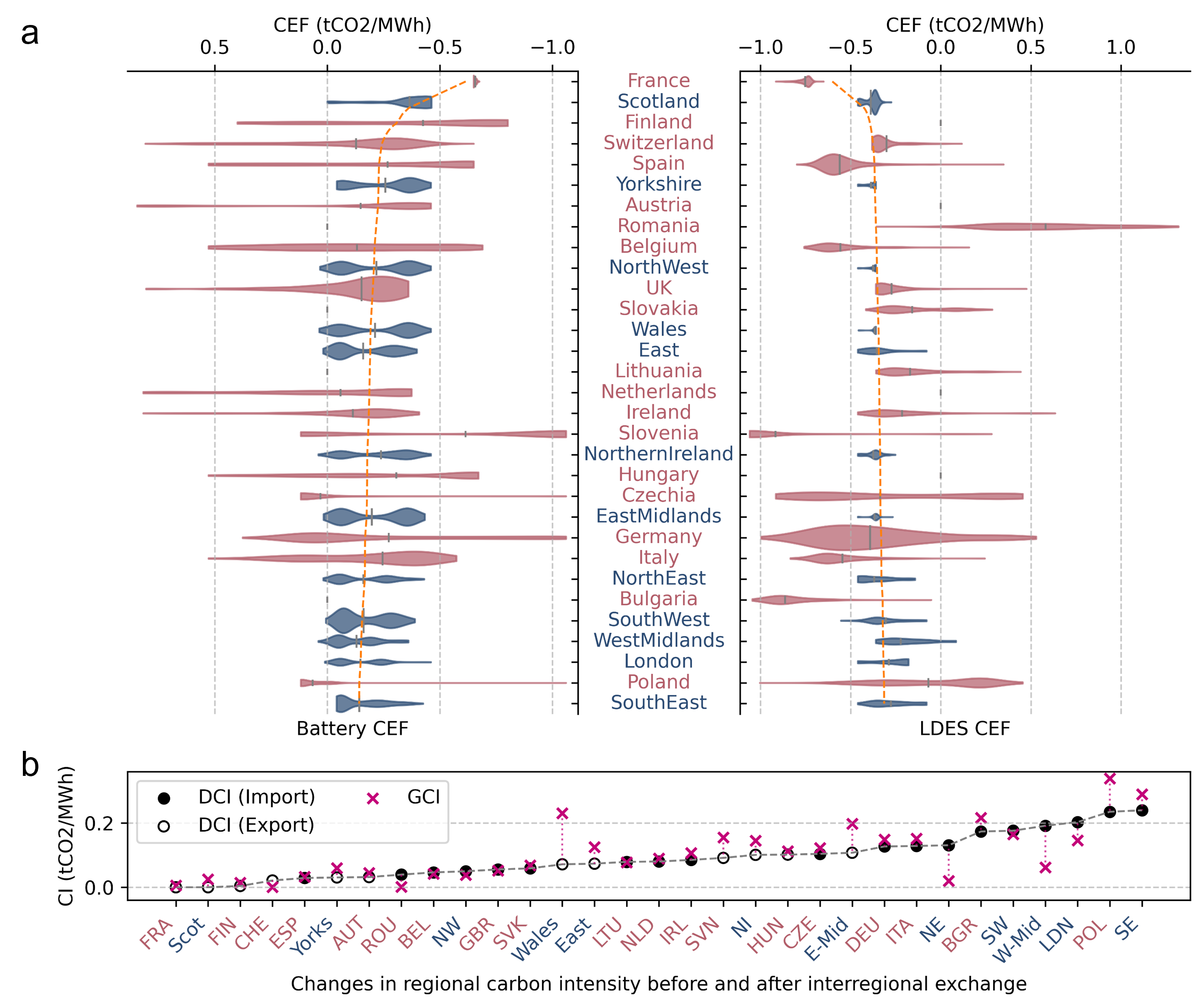}
    \caption{\textbf{Regional distributions of storage CEFs and changes in regional carbon intensity.} 
    \textbf{a,} Regional distributions of cycle emission factors for batteries and LDES across UK regions and EU countries. Battery CEFs are shown on the left and LDES CEFs on the right, with region names placed between the two distributions to enable direct comparison across storage types and regions. Regions are ordered by annual demand carbon intensity, and labels are colour-coded by case study (blue, UK; red, EU). Violin width is proportional to the total annual discharged energy of the corresponding storage type in each region, and the internal horizontal line marks the aggregated regional CEF. Negative CEF values indicate emission reductions. The two CEF axes are mirrored around the central region labels, so values closer to the centre indicate more negative CEFs and therefore stronger emission reductions for both storage types. Dashed curves show spline fits to the regional aggregated CEFs. The corresponding fitted equations are CEF = 0.0965 $\times$ log$_{10}$(DCI) $-$ 0.0811 and CEF = 0.0588 $\times$ log$_{10}$(DCI) $-$ 0.2767 for batteries and LDES, respectively. 
    \textbf{b,} Regional carbon intensity before and after interregional electricity exchange. The y axis shows carbon intensity (tCO$_2$/MWh). DCI denotes the average carbon intensity of electricity serving local demand after interregional exchanges, whereas GCI denotes the carbon intensity of local generation before exchanges. DCI (Import) refers to regions where annual net electricity imports exceed exports, whereas DCI (Export) refers to regions where annual net electricity exports exceed imports. Regions are aligned with \textbf{a}. Where the GCI point lies below the DCI curve, interregional energy exchange increases local carbon intensity; where it lies above the DCI curve, interregional energy exchange reduces local carbon intensity.}
    \label{fig:fig6}
\end{figure}

Interregional exchange substantially changes those opportunities \cite{Qiu2024,Brinkerink2024}. Regional electricity use is shaped by both local and imported supply rather than by local generation alone \cite{Zeyringer2018,Brinkerink2024} (Fig.~\ref{fig:fig5}a; Fig.~S8), so storage cannot be interpreted from local average carbon intensity alone \cite{Elenes2022,Steinsultz2024}. For both technologies, aggregated charging and discharging emission factors exceed the carbon intensity of electricity serving regional demand, showing that storage is governed by exchanged and marginal supply conditions. Batteries are more geographically dispersed and engage in denser cross-regional charging and discharging interactions, whereas LDES is concentrated in fewer, generally lower-carbon regions with more selective exchanges (Fig.~\ref{fig:fig5}b--e). Spatial explainability therefore requires tracing both sides of the cycle across the network: where charging electricity effectively comes from and which regional supply is displaced at discharge \cite{Qiu2024,Steinsultz2024}.

Under this accounting, LDES delivers more consistently favourable regional emissions outcomes than batteries in both the UK and EU, although the EU shows much greater spatial heterogeneity (Fig.~\ref{fig:fig6}a). In the UK, aggregated regional CEFs remain uniformly negative, whereas in the EU they span from strongly negative to positive values, reflecting broader variation in regional mixes and exchange patterns. Across this gradient, the fitted LDES curve remains more negative than the battery curve. At the cycle level, battery distributions are more often split between quadrants~3 and~4 (in an emission minimization case), and in some regions extend into quadrant~1, whereas LDES remains more concentrated in quadrant~4 (Figs.~S9-S11). Its spatial advantage therefore arises not simply from lower charging emissions, but from more reliable access to co-benefit cycles under regional exchange.

The spatial extension also identifies three locational regimes (Fig.~\ref{fig:fig6}b). Clean-surplus exporters, such as Scotland in the UK and France in the EU, combine cleaner charging with displacement of more carbon-intensive supply elsewhere in the network, so their cycles cluster in quadrant~4. Deficit importers, such as London and Romania, are more exposed to higher-carbon external supply and shift more readily toward quadrants~3 or~1 unless charging is steered toward cleaner periods. Ultra-high-emission peaker exporters, such as the SouthEast and Bulgaria, can still deliver strong abatement because discharge replaces very carbon-intensive marginal generation, but more often through quadrant~3 trade-offs and higher system cost. The most suitable storage locations are therefore not simply the cleanest regions, but those where clean charging coincides with strong displacement potential and a larger share of quadrant~4 cycles \cite{Brinkerink2024,Steinsultz2024}. This expands cycle-resolved explainability from identifying when storage creates value to identifying where network conditions allow that value to be realized. Together, the preceding sections resolve storage's public value across time (S2--S3), technology and duration (S4--S5), and space (S6), providing a basis for interpreting the gaps between current market arrangements and whole-system value.

\section*{Discussion and Conclusions}

These results reveal three distinct gaps between asset-level operation and whole-system value \cite{Beuse2021,Qin2023}. The first is an operational alignment gap: within today’s systems, some storage cycles are more strongly aligned with joint cost and emissions reduction than others, and improved incentives can steer storage towards those better opportunities \cite{Qin2023}. The second is a realization gap: present-day systems already contain long-duration co-benefit opportunities that would jointly reduce cost and emissions, but current market arrangements may not fully reward or verify these long-duration low-carbon flexibility regimes \cite{Sepulveda2021,Staadecker2024}. The third is a structural gap: in more carbon-intensive supply mixes, the stock of intrinsically low-cost, low-emission cycles remains limited \cite{Moradi2024,Victoria2022}. Operational reform can therefore improve alignment at the margin, but it cannot substitute for supply-side decarbonisation \cite{Kucuksayacigil2025,Moradi2024}.

This framing helps interpret current electricity markets \cite{Qin2023,BajoBuenestado2025}. Existing energy, balancing and ancillary-service arrangements already reward many frequent short-cycle actions because they align well with operational patterns of generators, helping explain the commercial visibility of lithium-ion batteries \cite{Jiang2025a,Kittner2017}. But the explainability framework shows that commercially rational battery dispatch can still remain on a cost-emission trade-off frontier from a whole-system perspective \cite{Beuse2021,Hittinger2015}. Current market signals therefore capture part of storage value, but not its full public value \cite{Qin2023}. Better short-run incentives may reduce this mismatch, especially by discouraging cycles that lower cost while increasing emissions, but they do not create co-benefit opportunities that the surrounding system does not yet provide \cite{Steinsultz2024,NaviaSimon2025}.

The case for LDES is different. Across modeled systems, LDES contributes more consistently through co-benefit cycles, especially when value depends on sustained charging from clean surplus and prolonged displacement of high-emission supply during stressed periods \cite{Sepulveda2021,Staadecker2024}. These operating regimes are not well represented by short-horizon arbitrage revenues or generic adequacy products \cite{Olsen2019}. A critical implication is therefore not simply that LDES has high modeled value, but that present-day market arrangements may under-reward operating regimes that could reduce both cost and emissions from a whole-system perspective \cite{Ferrara2019}. In many systems, analogous temporal functions are still provided by incumbent thermal flexibility, which may preserve adequacy but also preserves emissions exposure and leaves part of the system’s embedded co-benefit potential unrealized \cite{An2025,Guo2024}.

The implication is not that LDES should be rewarded unconditionally as a technology class. Rather, markets and procurement mechanisms should reward the operating regimes through which this underused co-benefit potential is created \cite{NaviaSimon2025}. These include sustained availability across prolonged stressed periods, discharge when system cost and emissions are jointly high, and charging linked to verified use of clean surplus \cite{Staadecker2024}. The framework provides a grounded basis for both ex-ante market design and ex-post verification by specifying the missing product as a recurring set of cycle properties that generate co-benefits \cite{Steinsultz2024}.

This study has several limitations. First, it isolates operational effects after storage assets are built. We quantify the system costs and emissions induced by storage dispatch, but do not include life-cycle impacts such as embodied emissions, manufacturing, end-of-life processes, or fixed non-energy operation and maintenance \cite{Paul2024,Le2024}. These factors are therefore not endogenized in siting, sizing or dispatch. Second, interregional transfers are represented through central dispatch with aggregate transmission limits. This captures broad network constraints but not nodal power-flow physics or local congestion, so the ease of rebalancing across regions may be overstated \cite{Brinkerink2024}. Third, the analysis contrasts stylized least-cost and emissions-minimizing dispatches to bound how operating priorities affect storage behaviour. These cases are diagnostically useful, but they do not directly model merchant bidding, realized revenues or detailed market-clearing rules \cite{DESNZ2024REMA}. The market-design implications should therefore be interpreted as insights from operation-to-system valuation rather than as direct simulations of market outcomes. Fourth, the analysis focuses on grid-connected, post-generation electricity storage. Fuel-side or pre-generation storage options may provide complementary flexibility, but their cost and emissions implications involve a different system boundary and are not compared here.

Overall, this paper shows that whole-system storage value can be resolved into identifiable operating cycles with distinct cost and CO$_2$ consequences \cite{Beuse2021}. The framework explains why batteries can be commercially attractive under short-cycle price signals while only partly aligned with whole-system decarbonisation value \cite{Jiang2025a,Beuse2021}, and why long-duration storage can provide substantial but currently underused opportunities to reduce both cost and emissions \cite{Sepulveda2021,Staadecker2024}. The main advance is therefore not only a new metric, but a practical bridge between system planning and storage operation that identifies which operating regimes create public value, and under what system conditions that value can be realized.

\section*{Method}

\subsection*{Model and data sources for the UK case}

The analysis presented in the UK case uses the PyPSA-UK electricity system optimization model \cite{Ember2022,Lyden2024}. PyPSA-UK is a Python-based open-source toolkit designed for simulating and optimizing modern electricity systems, and has been used in UK power-system research and scenario analysis \cite{Daggash2019,Lyden2024}. The model simulates each hour over a year, representing the UK electricity network through 12 regional nodes. Beyond the original generator dataset provided by PyPSA-UK, additional current and planned generation assets were integrated using established databases, including the Digest of UK Energy Statistics (DUKES) \cite{DESNZ2025DUKES} and the Renewable Energy Planning Database \cite{DESNZ2025REPD}. The installed capacities of generation and storage facilities, along with load data for System 2023 and System 2030, were reallocated based on the ``2023'' and ``2030 New Dispatch'' scenarios provided in the \textit{Clean Power 2030} report published by the National Energy System Operator (NESO) \cite{NESO2024}. Marginal generation costs for the two systems were both sourced from \textit{Electricity Generation Costs 2023} commissioned by the UK Department for Energy Security and Net Zero \cite{DESNZ2023GenCosts}. Emission factors were derived from the IPCC Emission Factor Database (2023), processed by Our World in Data \cite{OWIDEmissionsFactor}. The extension to the EU-wide cases is described in Supplementary Note S4. To reflect the operational characteristics of different storage types, the round-trip efficiency (RTE) of short-duration storage was set to 0.9 with the constraint that all stored energy must be discharged by the end of each day. In contrast, long-duration storage was assigned an RTE of 0.7, with no constraints on cycle duration. These two baseline storage types are stylized operational categories rather than exhaustive representations of all battery and LDES technologies. The purpose is to separate recurrent short-duration cycling from cross-timescale storage, while the sensitivity and technology-mapping analyses explore broader technology ranges. Comprehensive data details are accessible in the Data Availability section.

\subsection*{Model objectives}

The optimization model evaluates two primary objectives: minimizing total system operational costs and minimizing total system emissions \cite{Terlouw2023}. These are quantified as follows:

\begin{equation}
C_{\mathrm{total}} = \sum_{r}\sum_{u}\sum_{t} G_{r,u}(t)\,MC_u,
\label{eq:total-cost}
\end{equation}

\begin{equation}
Em_{\mathrm{total}} = \sum_{r}\sum_{u}\sum_{t} G_{r,u}(t)\,EF_u,
\label{eq:total-emissions}
\end{equation}

where $C_{\mathrm{total}}$ and $Em_{\mathrm{total}}$ represent the total cost and emissions of the power system, respectively. $G_{r,u}(t)$ represents the power output of generator $u$ in region $r$ at time $t$, while $MC_u$ denotes the marginal cost of generator $u$, and $EF_u$ denotes the emission factor of generator $u$.

Given potential identical marginal costs or emission factors among generators, secondary criteria are incorporated to maintain system emissions while minimizing costs:

\begin{equation}
\min C = C_{\mathrm{total}} + m\,Em_{\mathrm{total}},
\label{eq:min-cost-secondary}
\end{equation}

where $m$ is a small positive scalar. Analogously, the emissions minimization objective is

\begin{equation}
\min Em = Em_{\mathrm{total}} + m\,C_{\mathrm{total}}.
\label{eq:min-emissions-secondary}
\end{equation}

Since the objective function focuses on the operating costs and carbon emissions of generators, multiple storage dispatch solutions may achieve the same system-level optimum. Our subsequent analysis of storage-induced costs and emissions is therefore based on the specific dispatch solution obtained from the optimization.

\subsection*{Electricity system optimization results}

\setcounter{figure}{0}
\renewcommand{\thefigure}{M\arabic{figure}}
\renewcommand{\theHfigure}{M\arabic{figure}}

\begin{figure}[!htbp]
    \centering
    \includegraphics[width=\linewidth]{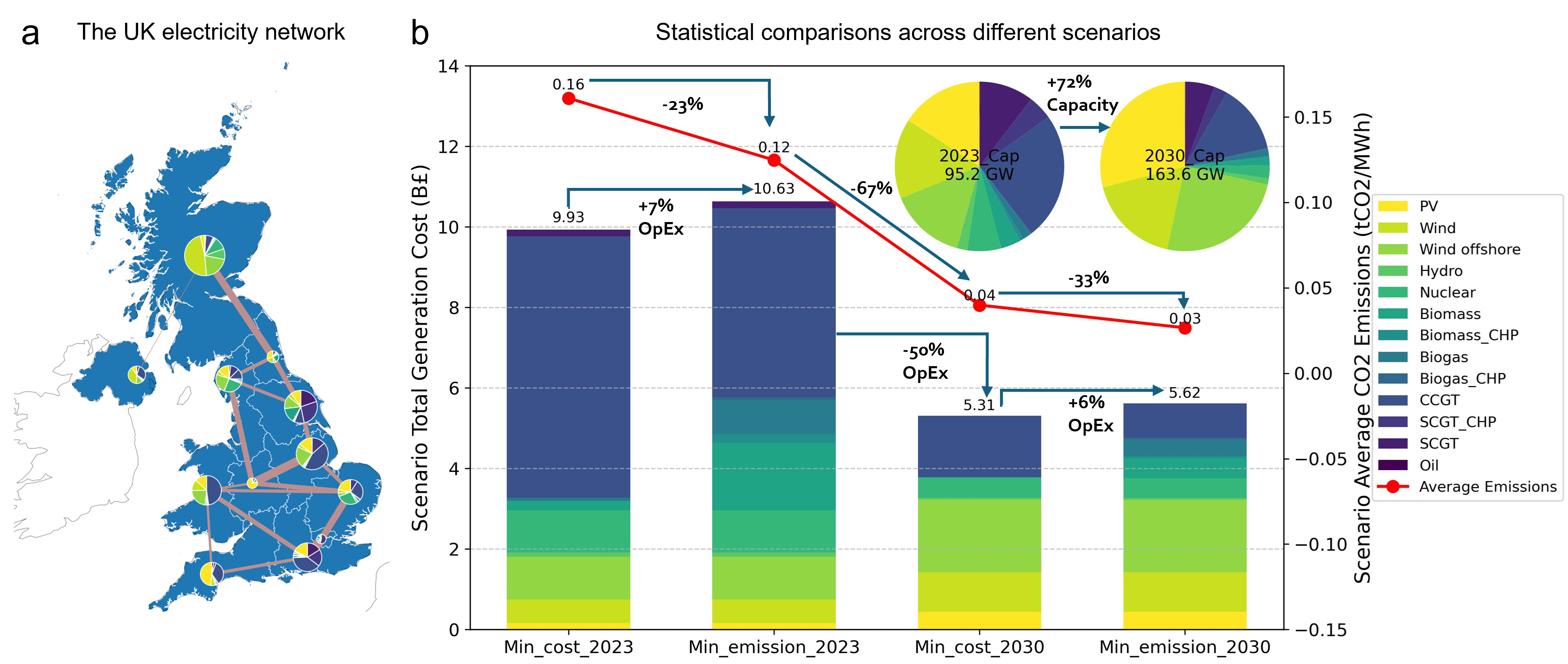}
    \caption{\textbf{Topology and statistical characterization of the UK electricity system.}
    \textbf{a,} The UK electricity network is segmented into 12 nodes, with their geographical distribution and interconnections shown explicitly. Line thickness denotes the transmission capacity between nodes, whereas node size reflects the installed generation capacity and the proportional composition of energy sources at each location.
    \textbf{b,} The bar chart shows the cost shares of generation technologies, as indicated in the legend; numbers above the bars denote total system operating costs in each scenario (left y axis, billion GBP). The line graph shows the corresponding average carbon emissions, calculated as total emissions divided by total system load (right y axis, tCO$_2$/MWh). The pie chart (top right) illustrates the installed capacity shares of different generation technologies. \textit{Min\_cost} and \textit{Min\_emission} denote system operation under cost-minimization and emission-minimization objectives, respectively. 2023 and 2030 refer to System 2023 and System 2030, respectively.}
    \label{fig:M1}
\end{figure}

Fig.~\ref{fig:M1} illustrates the topology of the UK electricity system and provides a statistical comparison across the scenarios analysed. The corresponding results for the EU case are presented in Supplementary Fig.~S12.

Fig.~\ref{fig:M1}b provides statistical insight into electricity systems with different generation mixes. The pie charts show that, from System 2023 to System 2030, total installed generation capacity increases from 95.2\,GW to 163.6\,GW, representing a 72\% increase. At the same time, the generation mix shifts markedly from predominantly carbon-intensive sources in 2023 to primarily zero-carbon sources in 2030. This transformation in both installed capacity and generation mix leads to substantial reductions in system-wide operating costs and emissions. Comparing the \textit{Min\_emission\_2023} and \textit{Min\_cost\_2030} scenarios, total system operating costs decline from £10.63 billion to £5.31 billion, while average carbon emissions decrease from 0.12 to 0.04\,tCO$_2$/MWh. However, comparing the two optimization objectives within the same system shows that lower emissions are consistently achieved at the expense of higher operating costs. These results describe system-level cost and carbon outcomes at the grid scale, but they do not yet reveal the precise operational role of storage or quantify its contribution to these aggregate changes \cite{Victoria2020,NaviaSimon2025}.

\begin{revblock}
For visualisation only, generation technologies are grouped by relative marginal cost and emission intensity. low-cost and low-emission generators include PV, onshore and offshore wind, hydro, and nuclear; medium-cost generation refers to CCGT; medium-emission generation includes biomass, biomass CHP, biogas, and biogas CHP; high-cost generation includes biomass, biomass CHP, biogas, biogas CHP, SCGT CHP, SCGT, and oil; and high-emission generation includes CCGT, SCGT CHP, SCGT, and oil. Since the emission profile of bioenergy can vary considerably across feedstocks, conversion pathways, and accounting assumptions, the medium-emission bioenergy group is used here mainly to represent a relatively lower-emission dispatchable generation option within the visualisation, rather than to imply a universal lifecycle emission ranking.
\end{revblock}
\subsection*{Storage cycle extraction method}

\noindent\textbf{1) Range extraction method}

The extraction procedure adapts rainflow counting to the storage state-of-charge trajectory. It decomposes the state-of-charge (SOC) series into symmetric charge–discharge segments, records the time points and energy shares assigned to each cycle, and removes zero-energy segments. This produces the cycle set used for CEF and CCF attribution. Consider an energy--time (SOC) curve $e$--$t$ with $N_e$ discrete points. Initially, an energy utilization sequence $Z = [1,1,\ldots,1]$ is defined over the full period for cycle extraction. All extrema and boundary points of the SOC curve are identified, forming a chronological time vector

\begin{equation}
\mathbf{t} = \begin{bmatrix} t_1 & t_2 & \ldots & t_{N_e} \end{bmatrix}.
\label{eq:time-vector}
\end{equation}

Here, $t_m$ denotes the time associated with the $m$-th extreme or boundary point. An initial range vector $L$ and corresponding start and end time vectors $A$ and $B$ are then computed \cite{Wang2022}:

\begin{equation}
\left\{
\begin{aligned}
l_m      &= \left| e(t_{m+1}) - e(t_m) \right|,\\
\alpha_m &= t_m,\\
\beta_m  &= t_{m+1},\\
L        &= \left[ l_1 \quad l_2 \quad \cdots \quad l_{N_e-1} \right],\\
A        &= \left[ \alpha_1 \quad \alpha_2 \quad \cdots \quad \alpha_{N_e-1} \right],\\
B        &= \left[ \beta_1 \quad \beta_2 \quad \cdots \quad \beta_{N_e-1} \right].
\end{aligned}
\right.
\label{eq:range-vectors}
\end{equation}

where $l_m$ is the magnitude of the $m$-th range, $\alpha_m$ and $\beta_m$ are the corresponding start and end points, and $e(t)$ denotes the SOC at time $t$.

\medskip
\noindent\textbf{2) Parameter extraction procedure}

\noindent\textit{Step 1. Initial check.} Select three consecutive ranges from $L$ and check whether $l_{m_2} \le l_{m_1}$. If the condition holds, proceed to Step 2; otherwise, iterate forward until the last range and then proceed to Step 4.

\medskip
\noindent\textit{Step 2. Secondary check.} Calculate the energy variation
\begin{equation}
d_n = e\bigl(t_{m_2}\bigr) - e\bigl(t_{m_1}\bigr).
\label{eq:dn-secondary}
\end{equation}
If $|d_n| > 0.1$, proceed to Step 3. Otherwise, merge ranges $m_1$ and $m_2$ into a new range $m_3$, remove ranges $m_1$ and $m_2$, update vector $L$, and return to Step 1.

\medskip
\noindent\textit{Step 3. Cycle extraction.} Record the cycle identifier $\lambda_n$, energy variation $d_n$, and cycle end time $\beta_n'$ as
\begin{equation}
\left\{
\begin{array}{l}
\lambda_n = n,\\[3pt]
d_n = e\bigl(t_{m_2}\bigr) - e\bigl(t_{m_2-1}\bigr),\\[3pt]
\beta_n' = \beta_{m_2}.
\end{array}
\right.
\label{eq:cycle-basic}
\end{equation}

Determine the cycle start time $\alpha_n'$. If $d_n > 0$, find the first time point $t$ before $\beta_n'$ satisfying $e(t) \le e(\beta_n')$, and record $t+1$ as $\alpha_n'$ for a charge-first cycle. If $d_n < 0$, identify the first time point $t$ before $\beta_n'$ satisfying $e(t) \ge e(\beta_n')$, and record $t+1$ as $\alpha_n'$ for a discharge-first cycle.

Record the energy proportion sequence up to the cycle endpoint and update the overall sequence $Z$:
\begin{equation}
\left\{
\begin{array}{l}
\tau_i = u_i,\\[3pt]
z_i = 0,
\end{array}
\right.
\qquad \alpha_n' + 1 \le i \le \beta_n'.
\label{eq:tau-z-update}
\end{equation}

To ensure charging--discharging symmetry, update the energy proportion at the start point:
\begin{equation}
\tau_{\alpha_n'} = \frac{e\bigl(\alpha_n' + 1\bigr) - e\bigl(\beta_n'\bigr)}{e\bigl(\alpha_n' + 1\bigr) - e\bigl(\alpha_n'\bigr)},
\label{eq:tau-start}
\end{equation}

\begin{equation}
z_{\alpha_n'} = z_{\alpha_n'} - \tau_{\alpha_n'}.
\label{eq:z-start}
\end{equation}

All time points where $\tau_i > 0$ in cycle $\lambda_n$ define the extracted cycle parameters:
\begin{equation}
\left\{
\begin{array}{l}
\lambda_n = n,\\[3pt]
T_n = \begin{bmatrix} \alpha_n' & \alpha_n' + 1 & \ldots & \beta_n' \end{bmatrix},\\[3pt]
\Gamma_n = \begin{bmatrix} \tau_{\alpha_n'} & \tau_{\alpha_n'+1} & \ldots & \tau_{\beta_n'} \end{bmatrix}.
\end{array}
\right.
\label{eq:cycle-parameters}
\end{equation}

Merge ranges $m_1$ and $m_2$ into a new range $m_3$, remove ranges $m_1$ and $m_2$, update vector $L$, and return to Step 1.

\medskip
\noindent\textit{Step 4. Final cycle extraction.} Extract the remaining time points $t_i$ and corresponding $\tau_i$ values where $z_i > 0$ from sequence $Z$ as the last cycle.

\medskip
\noindent\textit{Step 5. Parameter adjustment.} Exclude zero-energy points, removing associated time points and energy proportions, resulting in final cycles $\lambda_n$, corresponding time-point sequences $T_n$, and respective energy-proportion sequences $\Gamma_n$.

\subsection*{Guidelines for interregional energy interactions}

When allocating energy flows among regions with a focus on system-level carbon emissions, we adopt the principle ``local first, then external; clean first, then high-carbon.'' (If the objective is cost minimization rather than emissions, replace ``carbon'' with ``cost'' throughout.) In each region, energy supply may come from four sources---local generation, external generation, local storage discharge, and external storage discharge---and generation is categorized into four types by both location and carbon intensity (CI): local low-carbon, local high-carbon, external low-carbon, and external high-carbon. Demand consists of local load, local storage charging, and external storage charging. As an accounting convention, we assume that in each region all available energy is dispatched in the ordered sequence: local low-carbon generation, local high-carbon generation, local storage discharge, external low-carbon generation, external high-carbon generation, and external storage discharge, to satisfy in turn the aggregated demands (local load $+$ local charging $+$ external charging). Accordingly, the interregional energy--carbon accounting proceeds as follows.

\medskip
\noindent\textbf{1) Generation preprocessing}

\noindent\textit{Step 1.} Let $G_{r,u}(t)$ be the generation from resource $u$ in region $r$ at time $t$, and let $D_r(t)$ be the corresponding storage discharge. Denote the total load in region $r$ by $L_r(t)$ and the total storage charging by $C_r(t)$. Order all generation types $u = 1,\ldots,U$ in each region by ascending carbon emission. Define each region's net surplus and deficit as

\begin{equation}
g_r^{sp}(t) = \max\!\left(\sum_{u=1}^{U} G_{r,u}(t) + D_r(t) - L_r(t) - C_r(t),\,0\right),
\label{eq:regional-surplus}
\end{equation}

\begin{equation}
g_r^{st}(t) = \max\!\left(L_r(t) + C_r(t) - \sum_{u=1}^{U} G_{r,u}(t) - D_r(t),\,0\right).
\label{eq:regional-deficit}
\end{equation}

\noindent\textit{Step 2.} Within each surplus region $r$, remove exactly $g_r^{sp}(t)$ units from $G_{r,u}(t)$ by depleting the highest-carbon resources first. Denote the post-deduction values by

\begin{equation}
\widetilde{G}_{r,u}(t) = G_{r,u}(t) - \Delta_{r,u}(t),
\label{eq:post-deduction-generation}
\end{equation}

where $\sum_u \Delta_{r,u}(t) = g_r^{sp}(t)$ and the depletion order follows $u \in (U \rightarrow 1)$.

\noindent\textit{Step 3.} Identify all regions whose deficit $g_k^{st}(t) > 0$. Redistribute each $\Delta_{r,u}(t)$ to those deficit regions $k$ in proportion to their deficits:

\begin{equation}
\Delta_{r \rightarrow k,u}^{recv}(t) = \Delta_{r,u}(t)
\frac{g_k^{st}(t)}{\sum_{m: g_m^{st}(t) > 0} g_m^{st}(t)}.
\label{eq:received-generation}
\end{equation}

\noindent\textit{Step 4.} Add all such received amounts to each deficit region to obtain the interregional adjusted generation matrix:

\begin{equation}
\widehat{G}_{k,u}(t) = \widetilde{G}_{k,u}(t) + \sum_{r=1}^{R} \Delta_{r \rightarrow k,u}^{recv}(t).
\label{eq:adjusted-generation}
\end{equation}

\medskip
\noindent\textbf{2) Remaining-generation preprocessing}

\noindent\textit{Step 1.} Let $G_{r,u}^{norm}(t)$ be the maximum available capacity of resource $u$ in region $r$. Define the initial remaining generation matrix as

\begin{equation}
G_{r,u}^{rem}(t) = G_{r,u}^{norm}(t) - G_{r,u}(t).
\label{eq:remaining-generation}
\end{equation}

\noindent\textit{Step 2.} For each region $r$, identify its marginal resource index $u_r^{*}(t)$ by scanning resources in descending-emission order (highest-carbon first) and picking the first $u$ for which $G_{r,u}(t) > 0$. Zero out all lower-carbon types than this marginal resource, i.e.

\begin{equation}
G_{r,u}^{rem}(t) = 0,
\qquad \forall u: \mathrm{emission}(u) < \mathrm{emission}\bigl(u_r^{*}(t)\bigr).
\label{eq:remaining-generation-zero}
\end{equation}

\noindent\textit{Step 3.} Compute the remaining surplus and deficit as

\begin{equation}
g_r^{rem,sp}(t) = \max\!\left(\sum_{u} G_{r,u}^{rem}(t) - D_r(t),\,0\right),
\label{eq:remaining-surplus}
\end{equation}

\begin{equation}
g_r^{rem,st}(t) = \max\!\left(D_r(t) - \sum_{u} G_{r,u}^{rem}(t),\,0\right).
\label{eq:remaining-deficit}
\end{equation}

\noindent\textit{Step 4.} Within each surplus region $r$, remove exactly $g_r^{rem,sp}(t)$ units from $G_{r,u}^{rem}(t)$ by depleting the highest-carbon resources first:

\begin{equation}
\widetilde{G}_{r,u}^{rem}(t) = G_{r,u}^{rem}(t) - \Delta_{r,u}^{rem}(t),
\label{eq:remaining-post-deduction}
\end{equation}

where $\sum_u \Delta_{r,u}^{rem}(t) = g_r^{rem,sp}(t)$ and the depletion order follows $u \in (U \rightarrow 1)$.

\noindent\textit{Step 5.} Identify all regions whose deficit $g_k^{rem,st}(t) > 0$ and redistribute each $\Delta_{r,u}^{rem}(t)$ to those deficit regions in proportion to their deficits.

When calculating generation, the power-balance constraint always ensures that $\sum_r g_r^{sp}(t) = \sum_r g_r^{st}(t)$, but the procedure for computing remaining-generation surplus does not satisfy this equality. Therefore, the interregional transfer volumes are determined by the relative magnitudes of $\sum_r g_r^{rem,sp}(t)$ and $\sum_r g_r^{rem,st}(t)$, and each deficit region can only receive energy above its own marginal resources.

Define $g_k^{rem,st}(t\mid 1)$ as the shortage of resource $1$ in region $k$ at time $t$, and adopt the following base case:

\begin{equation}
g_k^{rem,st}(t\mid 1) = g_k^{rem,st}(t).
\label{eq:base-shortage}
\end{equation}

The set of regions that experience a shortage of resource $u$ at time $t$ is defined as follows.

\begin{equation}
\varepsilon_u(t) = \left\{ k : u_k^{*}(t) \le u \text{ and } g_k^{rem,st}(t\mid u) > 0 \right\}.
\label{eq:shortage-set}
\end{equation}

The total remaining shortage and total remaining surplus of resource $u$ at time $t$ are defined as follows.

\begin{equation}
g_u^{rem,st}(t) = \sum_{k \in \varepsilon_u(t)} g_k^{rem,st}(t\mid u),
\label{eq:total-remaining-shortage}
\end{equation}

\begin{equation}
g_u^{rem,sp}(t) = \sum_r \Delta_{r,u}^{rem}(t).
\label{eq:total-remaining-surplus}
\end{equation}

Under discharge-substitution transfers, the total transferable quantity of resource $u$ at time $t$ is given by the following expression.

\begin{equation}
\varsigma_u(t) = \min\!\left(g_u^{rem,st}(t),\,g_u^{rem,sp}(t)\right).
\label{eq:transferable-quantity}
\end{equation}

The source-side and destination-side allocation shares for transfers of resource $u$ at time $t$ are defined as follows.

\begin{equation}
\varphi_{r,u}(t)  =
\begin{cases}
\dfrac{\Delta_{r,u}^{rem}(t)}{g_u^{rem,sp}(t)}, & g_u^{rem,sp}(t) > 0,\\
0, & \text{otherwise}.
\end{cases}
\label{eq:source-share}
\end{equation}

\begin{equation}
\phi_{k,u}(t)  =
\begin{cases}
\dfrac{g_k^{rem,st}(t\mid u)}{g_u^{rem,st}(t)}, & k \in \varepsilon_u(t) \text{ and } g_u^{rem,st}(t) > 0,\\
0, & \text{otherwise}.
\end{cases}
\label{eq:destination-share}
\end{equation}

Accordingly, the quantity transferred from source region $r$ to destination region $k$ for resource $u$ at time $t$ is specified as follows.

\begin{equation}
\Delta_{r \rightarrow k,u}^{recv,rem}(t) = \varsigma_u(t)\,\varphi_{r,u}(t)\,\phi_{k,u}(t).
\label{eq:transfer-quantity}
\end{equation}

The shortage of resource $u+1$ in region $k$ at time $t$ is updated according to the following rule.

\begin{equation}
g_k^{rem,st}(t\mid u+1) =
\begin{cases}
\max\!\left(g_k^{rem,st}(t\mid u) - \sum_r \Delta_{r \rightarrow k,u}^{recv,rem}(t),\,0\right), & u \in (1 \rightarrow U),\\
g_k^{rem,st}(t\mid u), & \text{otherwise}.
\end{cases}
\label{eq:update-shortage}
\end{equation}

\noindent\textit{Step 6.} Add all such received amounts to each deficit region to obtain the interregional adjusted remaining-generation matrix:

\begin{equation}
\widehat{G}_{k,u}^{rem}(t) = \widetilde{G}_{k,u}^{rem}(t) + \sum_r \Delta_{r \rightarrow k,u}^{recv,rem}(t).
\label{eq:adjusted-remaining-generation}
\end{equation}

\medskip
\noindent\textbf{3) Aggregate national-level generation}

Sum each resource type $u$ over all regions to obtain the national generation and remaining generation.

\begin{equation}
\widehat{G}_{u}^{nation}(t) = \sum_{k=1}^{R} \widehat{G}_{k,u}(t),
\label{eq:national-generation}
\end{equation}

\begin{equation}
\widehat{G}_{u}^{rem,nation}(t) = \sum_{k=1}^{R} \widehat{G}_{k,u}^{rem}(t).
\label{eq:national-remaining-generation}
\end{equation}

\medskip
\noindent\textbf{4) Compute storage-cycling indicators}

Using the indicator definitions and calculation methods specified in the next subsection, $\widehat{G}_{u}^{nation}$ and $\widehat{G}_{u}^{rem,nation}$ are used to compute the national-level storage-cycling indicators, while $\widehat{G}_{k,u}$ and $\widehat{G}_{k,u}^{rem}$ are used to compute the regional storage-cycling indicators.

\subsection*{Factors definition and calculation method}

For energy storage, the carbon emissions and costs associated with its usage are closely related to the types of energy utilized during charging and discharging. The cycle factors for energy storage are defined as follows:

\begin{equation}
CEF_{s,\lambda} = \frac{Em_{charged,s,\lambda} - Em_{avoided,s,\lambda}}{E_{dis,s,\lambda}},
\label{eq:cef}
\end{equation}

\begin{equation}
CCF_{s,\lambda} = \frac{C_{charged,s,\lambda} - C_{avoided,s,\lambda}}{E_{dis,s,\lambda}},
\label{eq:ccf}
\end{equation}

\begin{equation}
ACEF_s = \frac{\sum_{\lambda} Em_{charged,s,\lambda} - \sum_{\lambda} Em_{avoided,s,\lambda}}{\sum_{\lambda} E_{dis,s,\lambda}},
\label{eq:acef}
\end{equation}

\begin{equation}
ACCF_s = \frac{\sum_{\lambda} C_{charged,s,\lambda} - \sum_{\lambda} C_{avoided,s,\lambda}}{\sum_{\lambda} E_{dis,s,\lambda}}.
\label{eq:accf}
\end{equation}

Here, $CEF_{s,\lambda}$ and $CCF_{s,\lambda}$ represent the cycle emission factor and cycle cost factor of storage technology $s$ in cycle $\lambda$, respectively; $ACEF_s$ and $ACCF_s$ represent the corresponding aggregated cycle emission factor and aggregated cycle cost factor; $Em_{charged,s,\lambda}$ and $C_{charged,s,\lambda}$ denote the carbon emissions and cost caused by charging storage $s$ in cycle $\lambda$; $Em_{avoided,s,\lambda}$ and $C_{avoided,s,\lambda}$ refer to the carbon emissions and cost avoided due to storage $s$ discharging in cycle $\lambda$; and $E_{dis,s,\lambda}$ represents the discharged energy of storage $s$ in cycle $\lambda$.

\paragraph{Charging process.}
Based on the current power generation, the generation types are ranked in descending order of their carbon emissions. The generation volumes are accumulated sequentially until the total satisfies the storage charging demand. The carbon emissions (cost) of the energy used in charging are calculated as the weighted sum of generation energy and the corresponding emission (cost) factors:

\begin{equation}
E_{cha,s,\lambda} = \sum_{t \in \lambda} E_{cha,s}(t),
\label{eq:echa-cycle}
\end{equation}

\begin{equation}
Em_{charged,s,\lambda} = \sum_{t \in \lambda} Em_{charged,s}(t),
\label{eq:emcharged-cycle}
\end{equation}

\begin{equation}
Em_{charged,s}(t) = \sum_u \widehat{G}_u(t)\,EF_u,
\label{eq:emcharged-time}
\end{equation}

\begin{equation}
C_{charged,s,\lambda} = \sum_{t \in \lambda} C_{charged,s}(t),
\label{eq:ccharged-cycle}
\end{equation}

\begin{equation}
C_{charged,s}(t) = \sum_u \widehat{G}_u(t)\,MC_u,
\label{eq:ccharged-time}
\end{equation}

where $\sum_u \widehat{G}_u(t) = \sum_s E_{cha,s}(t) = E_{cha}(t)$, with the accumulation order following $u \in (U \rightarrow 1)$, and $E_{cha,s,\lambda}$ denotes the energy charged by storage $s$ during cycle $\lambda$.

\paragraph{Discharging process.}
The remaining generation capacities of all generators at the current time are calculated. These capacities are ranked in ascending order of their carbon emissions. Starting from the first non-zero generation type, the remaining capacities are accumulated sequentially until the total satisfies the storage discharging demand. The avoided carbon emissions (cost) due to discharging are then calculated as the weighted sum of remaining generation capacities and their corresponding emission (cost) factors:

\begin{equation}
E_{dis,s,\lambda} = \sum_{t \in \lambda} E_{dis,s}(t),
\label{eq:edis-cycle}
\end{equation}

\begin{equation}
Em_{avoided,s,\lambda} = \sum_{t \in \lambda} Em_{avoided,s}(t),
\label{eq:emavoided-cycle}
\end{equation}

\begin{equation}
Em_{avoided,s}(t) = \sum_u \widehat{G}^{rem}_u(t)\,EF_u,
\label{eq:emavoided-time}
\end{equation}

\begin{equation}
C_{avoided,s,\lambda} = \sum_{t \in \lambda} C_{avoided,s}(t),
\label{eq:cavoided-cycle}
\end{equation}

\begin{equation}
C_{avoided,s}(t) = \sum_u \widehat{G}^{rem}_u(t)\,MC_u,
\label{eq:cavoided-time}
\end{equation}

where $\sum_u \widehat{G}^{rem}_u(t) = \sum_s E_{dis,s}(t) = E_{dis}(t)$, with the accumulation order following $u \in (1 \rightarrow U)$.

To compare the proposed metrics in this study with conventional measures such as MEF and AEF, we present the distribution of storage cycles under each metric across different scenarios in Supplementary Fig.~S13. Supplementary Fig.~S14 further illustrates the calculation outcomes of all three metrics for a representative storage cycle.

It is important to note that in an isolated system, the charging amount of energy storage is always less than the system's total generation capacity. However, discharged energy may exceed the system's remaining generation capacity. This ``excess'' discharging amount can be interpreted as a deficit in generation, requiring assignment of a suitable energy source to calculate the corresponding carbon emissions and cost. In this study, since CCGT is the primary dispatchable generator in the UK electricity system \cite{Daggash2019,DESNZ2025DUKES}, parameters for such additional energy requirements during discharge in the UK case are derived from CCGT. In the EU case, these parameters are based on the dominant dispatchable generation type in each country, defined as the technology with the largest installed capacity among dispatchable sources. For interconnected systems, both charging and discharging can interact with neighboring systems. Hence, during charging (discharging), carbon emissions and costs associated with energy exceeding local generation capacities are first calculated using energy emission and cost from (to) other regions at corresponding time points, while excess discharging parameters remain based on CCGT.

When multiple storage technologies coexist, the calculation sequence of different storage types may partially influence the resulting factor calculations. This study considers two storage types: battery and LDES. When both are utilized simultaneously, we consistently calculate battery first, followed by LDES. Supplementary Fig.~S15 compares the impact of this calculation order under the \textit{UK\_min\_emission\_2023} base scenario. The results show that the order of calculating different storage technologies affects only a small number of individual cycles and has no impact on the overall cycle distribution or aggregated performance metrics.

The charging and discharging factors for each storage cycle are further defined as follows:

\begin{equation}
\text{Charging cost factor}_{s,\lambda} = \frac{C_{charged,s,\lambda}}{E_{cha,s,\lambda}},
\label{eq:charging-cost-factor}
\end{equation}

\begin{equation}
\text{Discharging cost factor}_{s,\lambda} = \frac{C_{avoided,s,\lambda}}{E_{dis,s,\lambda}},
\label{eq:discharging-cost-factor}
\end{equation}

\begin{equation}
\text{Aggregated charging cost factor}_{s} = \frac{\sum_{\lambda} C_{charged,s,\lambda}}{\sum_{\lambda} E_{cha,s,\lambda}},
\label{eq:agg-charging-cost-factor}
\end{equation}

\begin{equation}
\text{Aggregated discharging cost factor}_{s} = \frac{\sum_{\lambda} C_{avoided,s,\lambda}}{\sum_{\lambda} E_{dis,s,\lambda}},
\label{eq:agg-discharging-cost-factor}
\end{equation}

\begin{equation}
\text{Aggregated charging emission factor}_{s} = \frac{\sum_{\lambda} Em_{charged,s,\lambda}}{\sum_{\lambda} E_{cha,s,\lambda}},
\label{eq:agg-charging-emission-factor}
\end{equation}

\begin{equation}
\text{Aggregated discharging emission factor}_{s} = \frac{\sum_{\lambda} Em_{avoided,s,\lambda}}{\sum_{\lambda} E_{dis,s,\lambda}},
\label{eq:agg-discharging-emission-factor}
\end{equation}

where $s$ denotes the storage technology and $\lambda$ represents a specific storage cycle.

\section*{Acknowledgements}

W.H. would like to acknowledge the support of UKRI-EPSRC [grant number EP/W027372/1].

\section*{Data and code availability}

The collected data and the code used to analyze the data in this study are available at \url{https://github.com/hui9883/CEF_2025}.

\bibliographystyle{unsrtnat}
\bibliography{references}

\end{document}